# BIG DATA IN CLOUD COMPUTING REVIEW AND OPPORTUNITIES

Manoj Muniswamaiah, Tilak Agerwala and  Charles Tappert

Seidenberg School of CSIS, Pace University, White Plains, New York

## ABSTRACT

*Big Data is used in decision making process to gain useful insights hidden in the data for business and engineering. At the same time it presents challenges in processing, cloud computing has helped in advancement of big data by providing computational, networking and storage capacity. This paper presents the review, opportunities and challenges of transforming big data using cloud computing resources.*

## 1. INTRODUCTION

The volume and information captured from various mobile devices and multimedia by organizations is increasing every moment and has almost doubled every year. This sheer volume of data generated can be categorized as structured or unstructured data that cannot be easily loaded into regular relational databases. This big data requires pre-processing to convert the raw data into clean data set and made feasible for analysis. Healthcare, finance, engineering, e-commerce and various scientific fields use these data for analysis and decision making. The advancement in data science, data storage and cloud computing has allowed for storage and mining of big data [1].

Cloud computing has resulted in increased parallel processing, scalability, accessibility, data security, virtualization of resources and integration with data storages. Cloud computing has eliminated the infrastructure cost required to invest in hardware, facilities, utilities or building large data centres. Cloud infrastructure scales on demand to support fluctuating workloads which has resulted in the scalability of data produced and consumed by the big data applications. Cloud virtualization can create virtual platform of server operating system and storage devices to spawn multiple machines at the same time. This provides a process to share resources and isolation of hardware to increase the access, management, analysis and computation of the data [2].

The main objective of this paper is to provide review, opportunities and challenges of big data applications in cloud computing which requires data to be processed efficiently and also provide some good design principles.

## 2. BIG DATA

Data which is huge, difficult to store, manage and analyse through traditional databases is termed as "Big Data". It requires a scalable architecture for their efficient storage, manipulation, and analysis. Such massive volume of data comes from myriad sources: smartphones and social

 

media posts; sensors, such as traffic signals and utility meters; point-of-sale terminals; consumer wearables such as fit meters and electronic health records. Various technologies are integrated to discover hidden values from these varied, complex data and transform it into actionable insight, improved decision making, and competitive advantage. The characteristics of big data are.

1.) Volume – Refers to incredible amount of data generated each second from different sources such as social media, cell phones, cars, credit cards, M2M sensors, photographs and videos which would allow users to data mine the hidden information and patterns found in them.

2.) Velocity - Refers to the speed at which data is being generated, transferred, collected and analysed. Data generated at an ever-accelerating pace must be analysed and the speed of transmission, and access to the data must remain instantaneous to allow for real-time access to different applications that are dependent on these data.

3.) Variety – Refers to data generated in different formats either in structured or unstructured format. Structured data such as name, phone number, address, financials, etc can be organized within the columns of a database. This type of data is relatively easy to enter, store, query, and analyse. Unstructured data which contributes to 80% of today's world data are more difficult to sort and extract value. Unstructured data include text messages, audio, blogs, photos, video sequences, social media updates, log files, machine and sensor data.

.

4.) Variability – Refers to the high inconsistency in data flow and its variation during peak period. The variability is due to multitude of data dimensions resulting from multiple disparate data types and sources. Variability can also refer to the inconsistent speed at which big data is ingested into the data stores.

5.) Value – Refers to the hidden value discovered from the data for decision making. Substantial value can be found in big data, including understanding your customers better, targeting them accordingly, optimizing processes, and improving machine or business performance.

6.) Veracity - Refers to the quality and reliability of the data source. Its importance is in the context and the meaning it adds to the analysis. Knowledge of the data's veracity in turn helps in better understanding the risks associated with analysis and business decisions based on data set.

7.) Validity - Refers to the accuracy of the data been collected for its intended use. Proper data governance practices need to be adopted to ensure consistent data quality, common definitions, and metadata.

8.) Vulnerability – Refers to the security aspects of the data been collected and stored.

9.) Volatility – Refers to how long data is valid and the duration for which it needs to be stored historically before it is considered irrelevant to the current analysis.

10.) Visualization – Refers to data being made understandable to nontechnical stakeholders and decision makers. Visualization is the creation of complex graphs that transforms the data into information, information into insight, insight into knowledge, and knowledge into advantage for decision making [3].

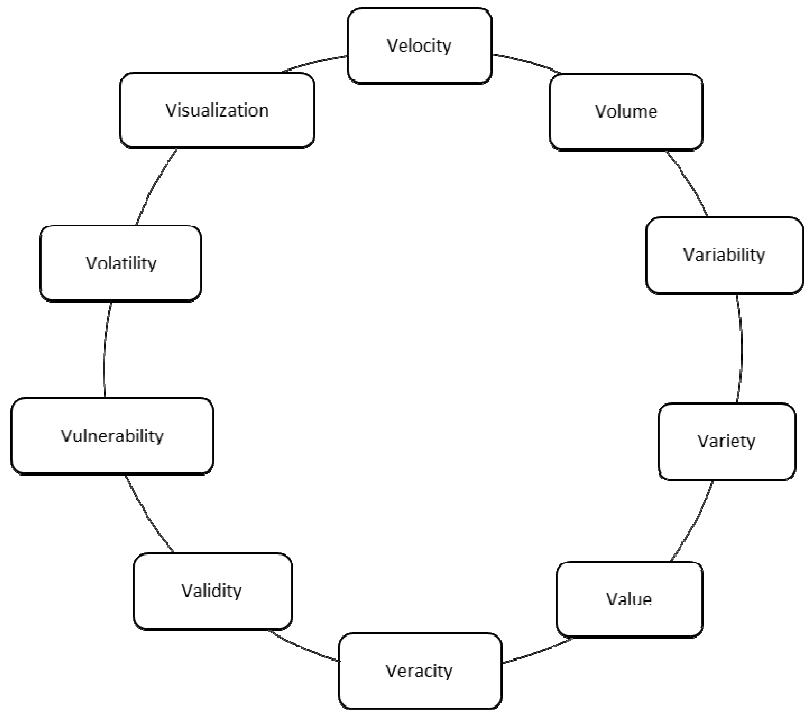

Figure 1: V's of Big Data

## 3. BIG DATA CLASSIFICATION

1. Analysis Type - Whether the data is analysed in real time or batch process. Banks use real time analysis for fraud detection whereas business strategic decisions can make use of batch process.

2. Processing Methodology - Business requirements determine whether predictive, ad-hoc or reporting methodology needs to be used.

3. Data Frequency - Determines how much of data is ingested and the rate of its arrival. Data could be continuous as in real-time feeds and also time series based.

4. Data Type - It could be historical, transactional and real-time such as streams.

5. Data Format - Structured data such as transactions can be stored in relational databases. Unstructured and semi-structured data can be stored in NoSQL data stores. Formats determine the kind of data stores to be used to store and process them.

6. Data Source - Determines from where the data is generated like social media, machines or human generated.

7. Data consumers - List of all users and applications which make use of the processed data [4].

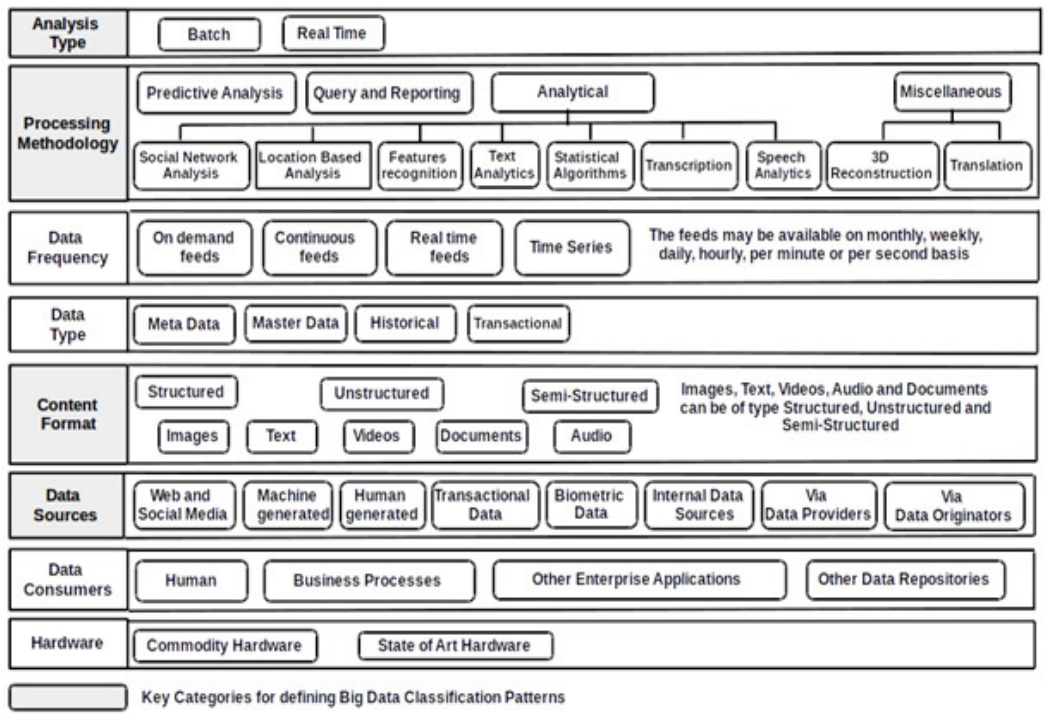

Figure 2: Big Data Classification

Big data is classified based upon its source, format, data store, frequency, processing methodology and analysis types as shown in Figure 2.

## 4. CLOUD COMPUTING

Cloud computing delivers computing services such as servers, storage, databases, networking, software, analytics and intelligence over the internet for faster innovation, flexible resources, heavy computation, parallel data processing and economies of scale. It empowers the organizations to concentrate on core business by completely abstracting computation, storage and network resources to workloads as needed and tap into an abundance of prebuilt services. Figure 3 shows the differences between on-premise and cloud services. It shows the services offered by each computing layer and differences between them [5].

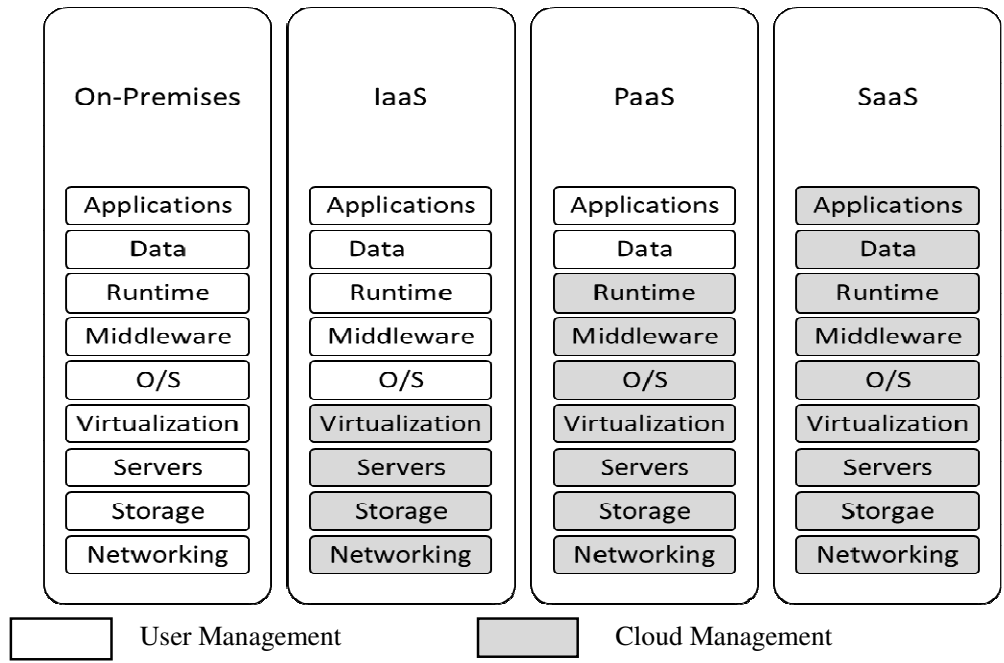

Figure 3: Summary of Key Differences

The array of available cloud computing services is vast, but most fall into one of the following categories:

*SaaS: Software as a Service*
Software as a service represents the largest cloud market and most commonly used business option in cloud services. SaaS delivers applications to the users over the internat. Applications that are delivered through SaaS are maintained by third-party vendors and interfaces are accessed by the client through the browser. Since most of SaaS applications run directly from a browser, it eliminates the need for the client to download or install any software. In SaaS vendor manages applications, runtime, data, middleware, OS, virtualization, servers, storage and networking which makes it easy for enterprises to streamline their maintenance and support.

*PaaS: Platform as a Service*
Platform as a Service model provides hardware and software tools over the internet which are used by developers to build customized applications. PaaS makes the development, testing and deployment of applications quick, simple and cost-effective. This model allows business to design and create applications that are integrated into PaaS software components while the enterprise operations or thirty-party providers manage OS, virtualization, servers, storages, networking and the PaaS software itself. These applications are scalable and highly available since they have cloud characteristics.

*IaaS: Infrastructure as a Service*
Infrastructure as a Service cloud computing model provides self-servicing platform for accessing, monitoring and managing remote data center infrastructures such as compute, storage and networking services to organizations through virtualization technology. IaaS users are responsible for managing applications, data, runtime, middleware, and OS while providers still manage virtualization, servers, hard drives, storage, and networking. IaaS provides same capabilities as data centers without having to maintain them physically [6]. Figure 4 represents the different cloud computing services been offered.

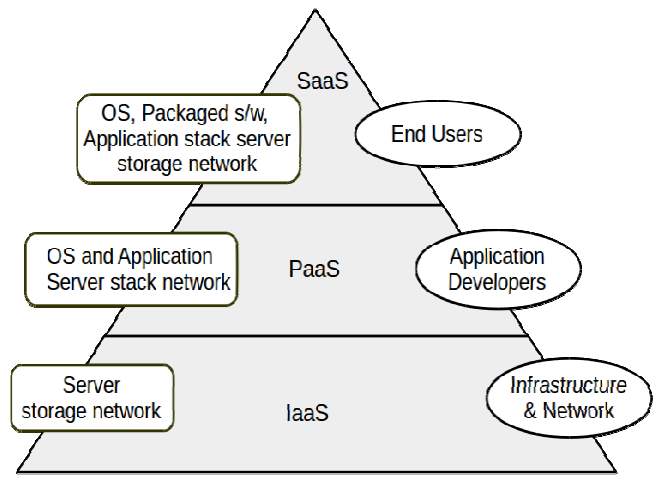

Figure 4: Primary Cloud Computing Services

## 5. RELATIONSHIP BETWEEN THE CLOUD AND BIG DATA

Big data and Cloud computing are closely related to each other. Big data is more about extracting value while cloud computing focuses on scalable, elastic, on-demand and pay-per-use self-service models. Big data requires massive on-demand computation power and distributed storage while cloud computing seamlessly provides elastic on-demand integrated computer resources, required storage and computing capacity to analyse big data. Cloud computing also offers the distributed processing for scalability and expansion through virtual machines to meet the requirements of exponential data growth. It has resulted in the expansion of analytical platforms that are designed to address the demands of users, especially large data-driven companies in providing contextual analysed data out of all the stored information. This has resulted in service providers like Amazon, Microsoft and Google in offering big data systems in cost efficient manner to capture data and adding analytics to offer proactive and contextual experiences.

Cloud computing environment is a new, great approach to providing IT-related services. Cloud computing environment has several providers and user terminals. It includes different kinds of software and hardware, pay-per-use or subscription-based services offered both through the internet and in real time. Data is collected using big data tools later it is stored and processed in cloud. Cloud provides on-demand resources and services for uninterrupted data management. The most common models for big data analytics are software services such as (SaaS), Platform service like (PaaS) and Infrastructure service like (IaaS). Recently Cloud analytics and Analytics as a Service (AaaS) are provided to clients on demand. Analytics as a Service (AaaS) provides subscription-based data analytics software and procedures through the cloud for a fast and scalable way to integrate data in semi-structured, unstructured and structured format, transform and analyse them.

Virtualization is the process of creating a software-based, or virtual representation of virtual desktop, servers, storage, operating system and network resources. It creates a virtual environment on an existing server to run desired applications, without interfering with any of the other services provided by the server or host platform to other users. It is a technique, which allows to share a single physical instance of a resource or an application among multiple customers and organizations. It does by assigning a logical name to a physical storage and providing a pointer to that physical resource when demanded. Virtualization enables multiple workloads, running on multiple physical servers, to be consolidated to run on individual virtual machines that are hosted on a single physical server. Minimizing the number of physical servers reduces data centre complexity, capital and administrative costs. This reduces and improves resource utilization and power consumption as compared to the multi-node setup. Virtual machines are highly flexible and portable which enables the workloads to be transferred to other physical servers during unplanned downtime, equipment or application maintenance. Virtualized big data applications like Hadoop provide benefits which cannot be provided using physical infrastructure in terms of resources utilization, cost and data management. Virtual data includes wide range of data sources and improves the data access from heterogeneous environments. It also enables high-speed data flow over the network for faster data processing.

Information privacy and security are one of the important aspects of big data. In cloud as data is hosted and processed on the third-party services and infrastructure, maintaining the privacy and security is a big challenge. Bigger the volume, variety, veracity of big data the higher the risk associated with its privacy and security. Mobile health has revolutionized health care delivery in many ways. It has enabled people to maintain their lifestyle, health and fitness, drug reference and diagnosis through various health care mobile applications. Since everything gets transferred over the mobile internet, it could pose a major threat to data privacy if the network is not secured properly or is exposed to intruders. Various new laws and privacy policies, regulations, safeguards, industry standards and Service level agreements must be maintained between providers and consumers in order to bring confidence in users and made sure that their data is not exposed. Security of big data in the cloud is important because data needs to be protected from malicious intruders treats and how the cloud providers securely maintain huge disk space [7].

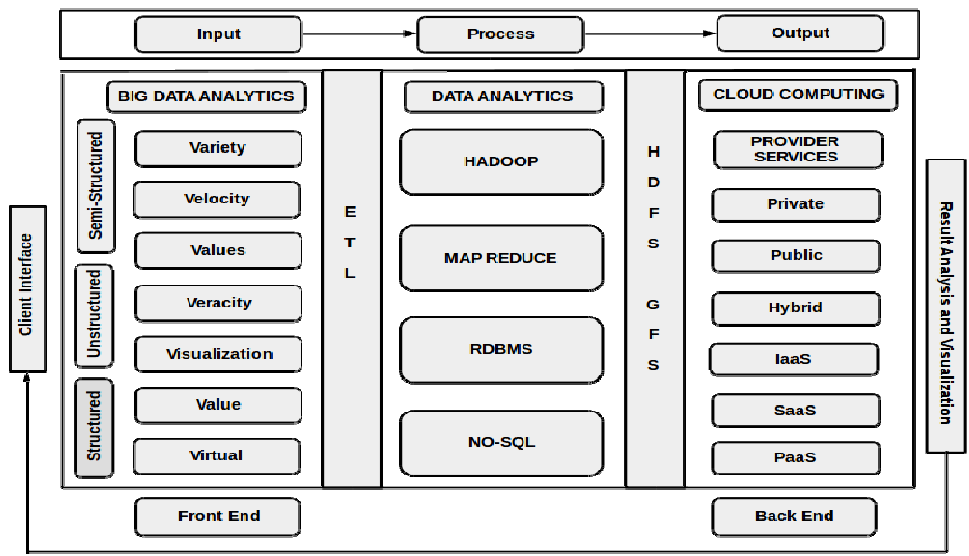

Figure 5: Big Data and Cloud Computing

The relationship between big data and cloud computing follows input, processing and output model as shown in Figure 5. The input is the big data obtained from various data sources such as cellular and other smart devices in either structured, unstructured or semi-structured format. This voluminous data is cleaned and then stored using Hadoop or other data stores. The stored data is in turn processed through cloud computing tools and techniques for providing services. Processing steps includes all the tasks required to transform input data. Output represents the value obtained after data is being processed for analysis and visualization [8].

Internet of Things (IoT) is one of the common factors between Cloud computing and big data. Data generated from IoT devices is massive and needs to be analysed in real time. Cloud providers allow data to be transmitted over internet or via lease lines and maintain them in data stores. Cloud computing techniques and tools are then used on the big data stored in cloud for filtering and analysis. It provides a pathway for the data to navigate, store and be analyzed. Cloud computing provides common platform for IoT and big data. Cloud computing, IoT and Big data have complementary relationship. IoT is the source of the data and big data is an analytical technology platform of the data as depicted in the Figure [6].

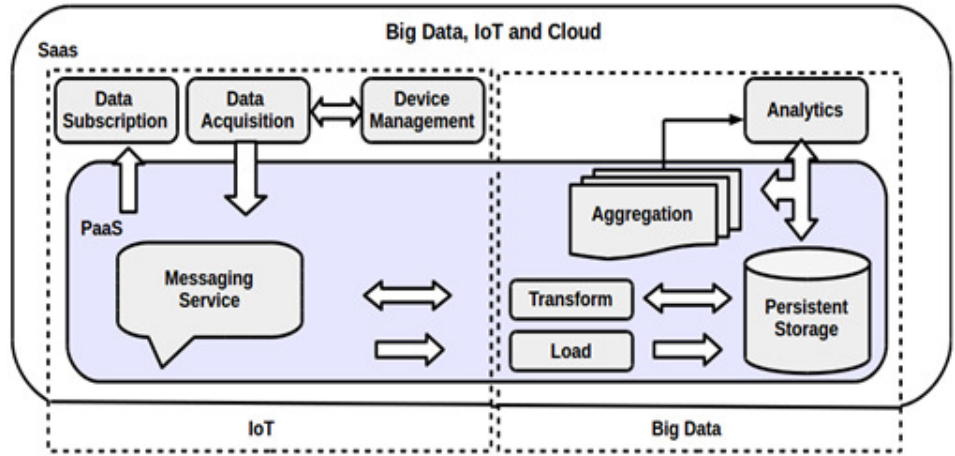

Figure 6: Overview of IoT, Big Data processing and Cloud Computing

## 6. CASE STUDIES

There are several case studies of big data on cloud computing.

### A. Tweet Mining in Cloud

Noordhuis et al. [9] used cloud computing to gather and analyse tweets. Amazon cloud infrastructure was used to perform all the computations. Tweets were crawled and later page ranking algorithm was applied. Page Ranking is used by Google to define the importance of a web page. On an overview, in Page Ranking Algorithm, if an author of a Web page A links to another Web page B, then author of Page A is likely to find Page B important. This way, the number of in-edges determine the importance of a certain page. Hence, more the number of in-links to a page, the more its importance. The same concept is applied to tweets, where instead of a web page, a tweet user following the other tweet is considered important. We need to crawl the Twitter social graph to compute PageRank. The data crawled had nearly 50 million nodes and 1.8 billion edges which is about two-thirds of the Twitter's estimated user base.

### B. Redbus

Redbus is an online travel agency for bus ticket booking in India. Redbus decided to use Google data infrastructure for data processing and analysis in order to improve customer sales and management of the ticket booking system. Google BigQuery enabled RedBus to process massive amounts of booking and inventory data within seconds. Applications that reside on multiple servers continuously streams customer searches, seat inventory and booking information to centralized data collection system. This massive data collected is then sent to BigQuery, which runs complex queries and within seconds provide answers to various analytical queries such as how many times a customer searched for a destination and found very few bus services available, any technical issues that might arise during booking and notify the necessary team. This whole infrastructure helped RedBus to fix glitches quickly, minimize lost sales and improve customer service. [9].

*C. Nokia mobile company*

Nokia mobile phones are been used by many people for telecommunication. In order to better understand their user interactions and improve the user experience with their phones, Nokia gathers large amount of data from mobile phones in petabyte scale. For deriving business decision strategies and get a holistic view of user interaction, the company implemented Hadoop data warehouse as an infrastructure that could store this daily petabyte of unstructured data collected from mobile phones in use, services, log files, and other sources. With this setup, Nokia was also able to perform complex processing and computation on their massive data and gain more analytical insights on their user interactions such as 'Which feature did they go to after this one?' and 'Where did they seem to get lost? It has also enabled them to create wide variety of applications such as 3D digital maps that incorporate traffic models that understand speed categories, recent speeds on roads, historical traffic models, elevation, ongoing events, video streams of the world, and more. [9].

# 7. DATA STORES

Modern databases need to handle large volume and different variety of data formats. They are expected to deliver extreme performance, provide high input/output operations per second required to deliver data to analytics tools and scale both horizontally and vertically to handle the dynamic data growth. Database architects have produced NoSQL and NewSQL as alternatives to relational database. NoSQL supports structured, semi-structured and unstructured data. It scales horizontally and provides BASE transactions. NewSQL is a new approach to relational databases that combines ACID transactions of RDBMSs and horizontal scalability of NoSQL. Below are characteristics of relational database, NoSQL and NewSQL [10, 11].

| Characteristics of Databases | Relational Database | NoSQL Database | New SQL Database |
|---|---|---|---|
| ACID property | ✓ | ✗ | ✓ |
| Analytical and OLTP support | ✓ | ✗ | ✓ |
| Data analysis | ✓ | ✗ | ✓ |
| Requires Schema | ✓ | ✗ | ✗ |
| Data format support | ✗ | ✓ | ✗ |
| Distributed parallel processing | ✓ | ✓ | ✓ |
| Scalability | ✗ | ✓ | ✓ |

# 8. HADOOP TOOLS AND TECHNIQUES

Big data applications use various tools and techniques for processing and analyses of the data. Below table represents some of them [12, 13].

| Tools/Techniques | Description | Developed by | Written in |
|---|---|---|---|
| HDFS | Redundant and Reliable massive data storage | Introduced by Google | Java |
| Map Reduce | Distributed data processing framework | Introduced by Google | Java |
| YARN | Cluster resource management framework | Apache | Java |
| Storm | Stream based task parallelism | Twitter | Clojure |
| Spark | Stream based data parallelism | Berkeley | Scala |
| Map Reduce | Java API. | Introduced by Google | Java |
| Pig | Framework to run script language Pig Latin | Yahoo | Java |
| Hive | SQL-like language  HiveQL | Facebook | Java |
| HCatalog | Relational table view of data in HDFS | Apache | Java |
| HBase | NoSQL column oriented | Google's BigTable | Java |
| Casandra | NoSQL column oriented | Facebook | Java |
| Flume | Import/Export unstructure or semi-structure data into HDFS. Data ingestion into HDFS. | Apache | Java |
| Sqoop | Tool designed for efficiently transferring bulk structured data (RDBMS) into HDFS and vies versa. | Apache | Java |
| Kafka | Distributed publish-subscribe messaging system for data integration | LinkedIn | Scala |
| Ambari | Web based cluster management UI | Hortonworks | Java |
| Mahout | Library of machine learning algorithms | Apache | Java |
| Oozie | Define collection of jobs with their execution sequence and schedule time | Apache | Java |
| Sentry | Role based authorization of data stored on an Apache Hadoop cluster. | Cloudera | Java |
| Zookeeper | Coordination service between hadoop ecosystems. | Yahoo | Java |

## 9. RESEARCH CHALLENGES

Big data are huge data sets that are very complex. The data generated is highly dynamic and this further adds to its complexity. The raw data must be processed in order to extract value from it. This gives rise to challenges in processing big data and business issues associated with it. Volume of the data generated worldwide is growing exponentially. Almost all the industries such as healthcare, automobile, financial, transportation etc rely on this data for improving their business and strategies. For example, Airlines does millions of transactions per day and have established data warehouses to store data to take advantages of machine learning techniques to get the insight of data which would help in the business strategies. Public administration sector also uses information patterns from data generated from different age levels of population to increase the productivity. Also, many of the scientific fields have become data driven and probe into the knowledge discovered from these data.

Cloud computing has been used as a standard solution for handling and processing big data. Despite all the advantages of integration between big data and cloud computing, there are several challenges in data transmission, data storage, data transformation, data quality, privacy, governance [14, 15].

*Data Transmission*
Data sets are growing exponentially. Along with the size, the frequency at which these real-time data are transmitted over the communication networks has also increased. Healthcare professions exchange health information such as high-definition medical images that are transmitted electronically while some of the scientific applications may have to transmit terabytes of data files that may take longer to traverse the network. In case of streaming applications, the correct sequence of the actual data packets is as critical as the transmission speed. Cloud data stores are used for data storage however, network bandwidth, latency, throughput and security poses challenges.

*Data Acquisition and Storage*
Data acquisition is the process of collecting data from disparate sources, filtering, and cleansing data before it can be stored in any data warehouses or storage systems. While acquiring big data, the main characteristics that pose a challenge are the sheer volume, greater velocity, variety of the big data. This demands more adaptable gathering, filtering, and cleaning algorithms that ensure that data are acquired in more time-efficient manner.

Data once acquired, needs to be stored in big capacity data stores which must provide access to these data in a reliable way. Currently there are Direct Attached Storage (DAS) and Network Attached Storage (NAS) storage technologies.

*Data Curation*
It refers to the active and ongoing management of data through its entire lifecycle from creation or ingestion to when it is archived or becomes obsolete and is deleted. During this process, data passes through various phases of transformation to ensure that data is securely stored and is retrievable. Organizations must invest in right people and provide them with right tools to curate data. Such an investment in the data curation will lead to better quantification of high-quality data.

*Scalability*
Scalability refers to the ability to provide resources to meet business needs in an appropriate way. It is a planned level of capacity that can grow as needed. It is mainly manual and is static. Most of the big data systems must be elastic to handle data changes. At the platform level there is vertical and horizontal scalability. As the number of cloud users and data increases rapidly, it becomes a challenge to exponentially scale the cloud's ability in order to provide storage and process too many individuals who are connected to the cloud at the same time.

*Elasticity*

It refers to the cloud's ability to reduce operational cost while ensuring optimal performance regardless of computational workloads. Elasticity accommodates to data load variations using replication, migration and resizing techniques all in a real-time without service disruption. Most of these are manual instead being automated.

*Availability*

Availability refers to on demand availability of the systems to authorized users. One of the key aspects of cloud providers is to allow users to access one or more data services in short time. As the business models evolve, it would lead to rising demand for more real time system availability.

*Data integrity*

Data Integrity refers to modification of data only by the authorized user in order to prevent misuse. Cloud based applications does allow its users to store and manage their data in cloud data centres, however these applications must maintain data integrity. Since the users may not be able to physically access the data, the cloud should provide mechanisms to check for the integrity of data.

*Security and Privacy*

Maintaining the security of the data stored in the cloud is very important. Sensitive and personal information that is kept in the cloud should be defined as being for internal use only, not to be shared with third parties. This would be a major concern when providing personalized and location-based services as access to personal information are required to produce relevant results. Each operation such as transmitting data over network, interconnecting the systems over network or mapping virtual machines to their respective physical machines must be done in a secured way.

*Heterogeneity*

Big data is vast and diverse. Cloud computing systems need to deal with different formats structures, semi-structured and unstructured data coming from various sources. Documents, photos, audio, videos and other unstructured data can be difficult to search and analyse. Having to combine all the unstructured data and reconcile it so that it can be used to create reports can be incredibly difficult in real time.

*Data Governance and Compliance*

Data governance specifies the exercise of control and authority over the way data needs to be handled and accountabilities of individuals when achieving business objectives. Data policies must be defined on the data format that needs to be stored, different constraint models that limits the access to underlying data. Defining the stable data policies in the face of increasing data size and demand for faster and better data management technology is not an easy task and its policies could lead to counter productiveness.

*Data Uploading*

It refers to the ease with which massive data sets can be uploaded to the cloud. Data is usually been uploaded through internet. The speed at which data is uploaded in turn depends network bandwidth and security. This calls for improvement and efficient data uploading algorithms to minimize upload times and provide a secure way to transfer data onto the cloud.

*Data Recovery*

It refers to the procedures and techniques by which the data can be reverted to its original state in scenarios such as data loss due to corruption or virus attack. Since periodic backups of petabytes of data is time consuming and costlier, it is necessary to identify a subset of data valuable to the organization for backup. If this subset of data is lost or corrupted, it take weeks to rebuild the lost data at these huge scales and result in more downtime for the users.

*Data Visualization*

Data Visualization is a quick and easy way to represent complex things graphically for better intuition and understanding. It needs to recognize various patterns and correlations hidden under massive data. Structured data can be represented in traditional graphical ways, whereas it is difficult to visualize high diversity, uncertain semi-structured and unstructured big data in real-time. In order to cope with such large dynamic data, there is a need for immense parallelization which is a challenge in visualization [16].

## 10. BIG DATA BUSINESS CHALLENEGES

*Utilities: Power consumption prediction*

Utility companies use smart meter to measure gas and electricity consumption. These devices generate huge volumes of data. A big data infrastructure needs to monitor and analyse power generation and consumption using smart meters.

*Social Network: Sentiment analysis*

Social networking companies such as Twitter needs to determine what users are saying and topics which are trending in order to perform sentiment analysis.

*Telecommunication: Predictive analytics*

Telecommunication provides need to build churn models which depends on the customer profile data attributes. Predictive analytics can predict churn by analysing the subscribers calling patterns.

*Customer Service: Call monitor*

Call center big data solutions use application logs to improve performance. The log files needs to be consolidated from different formats before they can be used for analysis.

*Banking: Fraud Detection*

Banking companies should be able to prevent fraud on a transaction or a user account. Big data solutions should analyse transactions in real time and provide recommendations for immediate action and stop fraud.

*Retailers: Product recommendation*

Retailers can monitor user browsing patterns and history of products purchased and provide a solution to recommend products based on it. Retailers need to make privacy disclosures to the users before implementing these applications [4].

## 11. GOOD PRINCIPLES

Below are some of the good design principles for big data applications

*Good Architectural Design*

Big data architecture should provide distributed and parallel processing through cloud services. Each node in the cluster must be independent, equal and parallel to other nodes. There must be no resources shared among these nodes as it would drastically reduce the scalability. NoSQL can be used for high performance and faster retrieval of data. Lambda and Kappa architectures can be used for processing in real-time and batch processing mode.

*Different Analytical Methods*

After collecting data from disparate sources, organizations must decide on the analytical method that best suits their business interest. Some of the commonly used analytical methods are predictive analytics, stream analytics, data preparation, fraud detection, sentiment analysis. Once, the analytical methods are decided, big data applications need to take the advantage of data mining, machine learning, distributed programming, statistical analysis, in-memory analytics and visualization techniques offered through cloud.

*Use appropriate technique*

No one technique can be used to analyse data. Since, most data that flows through is unstructured, we must use appropriate technology stack to analyse this complex data. For instance, if there is a need for real-time processing, Apache Spark can be used or if we need batch-processing, then Hadoop can be used. If data must be stored over multiple data centres, then geographic databases must be considered. Specific analytical tools that work well with established databases must be used.

*Use in-memory analytics*

In-memory database analytics can be used to execute analytics where data resides. It performs all needed analytical functions at runtime. Since the queries and the data reside in the same server's memory it reduces response time.

*Distributed data storage for in-memory analytics*

The data needs to be partitioned and stored in distributed data stores to take the advantage of in-memory analytics. Cloud computing infrastructure offers this distributed data storage solutions which must be adopted.

*Coordination between tasks and data is required*

To achieve scalability and fault-tolerance coordination between data and its processing tasks is required. Specialized cluster management frameworks such as Zookeeper can be used [16].

## 12. CONCLUSION

In the big data era of innovation and competition driven by advancements in cloud computing has resulted in discovering hidden knowledge from the data. In this paper we have given an overview of big data applications in cloud computing and its challenges in storing, transformation, processing data and some good design principles which could lead to further research.